\documentclass[12pt]{iopart}
\usepackage{graphicx}
\usepackage{amssymb,amsbsy}
\usepackage[mathscr]{eucal}
\begin{document}

\title{Thermodynamical instability of black holes}
\author{V.V. Kiselev}
\address{Russian State Research Center Institute for High Energy Physics,\\
Pobeda 1, Protvino, Moscow Region, 142281, Russia}
\address{Department of Theoretical Physics,\\
Moscow Institute of Physics and Technology (State University),\\
Institutsky 9, Dolgoprudny, Moscow Region, 141701,  Russia}
\eads{\mailto{Valery.Kiselev@ihep.ru}, 
}

\begin{abstract}
In contrast to Hawking radiation of black hole with a given spacetime
structure, we consider a competitive transition due to a heat transfer from a
hotter inner horizon to a colder outer horizon of Kerr black hole, that
results in a stable thermodynamical state of extremal black hole. In this
process, by supposing an emission of gravitational quanta, we calculate the
mass of extremal black hole in the final state of transition.
\end{abstract}


\maketitle

\section{Introduction} The black hole radiation, discovered by S.~Hawking
\cite{Hawking,Hawking2}, deals with the quantum field theory in curved
spacetime \cite{BirrelDavies}, which is held adiabatically constant, while
matter particles are created from virtual pairs due to the action of
gravitational force near the black hole horizon, that separates the particle
from its antiparticle of virtual pair, when particles outside the horizon
form the almost black body radiation spectrum, for the distant observer. From
the quantum theory point of view, the Hawking radiation corresponds to
diagonal matrix element of transition for the gravitational degrees of
freedom, since the spacetime structure is not changed, and particles are
created in the constant gravitational background.

However, we could challenge to investigate non-diagonal transitions between
gravitational states, that should change the spacetime structure, of course.
Such the quantum processes would be inherently different from the Hawking
radiation. But a strict consideration should rigorously involve a true theory
of quantum gravity, which is still not at our hands now (though see
\cite{LQG}). Therefore, we cannot calculate probabilities of non-diagonal
gravitational transitions between black holes.

Nevertheless, even in the classical formulation, the gravity establishes the
thermodynamical conception of black holes
\cite{Hawking,Hawking2,Bekenstein,Jacobson,Padmanabhan} regarding the
horizons in terms of entropy, temperature etc. In this way, the temperature
of black hole horizon $T_\textsc{bh}$ equals the temperature of Hawking
radiation, while the energy conservation dictates to introduce the entropy
$S_\textsc{bh}$, ascribed to the horizon because of the thermodynamical
relation between the variations $\delta M_\textsc{bh}=T_\textsc{bh}\delta
S_\textsc{bh}$, wherein the entropy obeys the connection to the horizon area
$\mathcal{A}_\textsc{bh}$,
\begin{equation}\label{SA}
    S_\textsc{bh}=\frac{1}{4G}\,\mathcal{A}_\textsc{bh},
\end{equation}
with $G$ being the Newton constant\footnote{In the text below we put $G=1$.},
while the temperature is given by the surface gravity, i.e. the acceleration
of free falling at the horizon $\kappa$,
\begin{equation}\label{kappa}
    T_\textsc{bh}=\frac{\kappa}{2\pi}.
\end{equation}
To concretize, the rotating black hole is described by the Kerr solution of
Einstein equations. Then, it has two horizons, the inner and outer ones, with
appropriate temperatures,
\begin{equation}\label{T+-}
    T_\pm=\frac{r_+-r_-}{\mathcal A_\pm},
\end{equation}
with relevant areas $\mathcal{A}_\pm$ at radiuses $r_+\geqslant r_-$. There
are two limits:
\begin{enumerate}
  \item at $r_-\to 0$ one gets the Schwarzschild black hole with zero
      orbital momentum and zero area of inner horizon $\mathcal A_-\to 0$
      giving the infinitely hot inner horizon contracted to the
      singularity at $r=0$;
  \item at $r_-\to r_+$ one gets the extremal black hole, which horizons
      are degenerated at absolute zero of temperature.
\end{enumerate}
Then, holding the constant spacetime structure, we require the inner and
outer spatial thermostats outside the horizons to be almost unchanged. It
means that virtual particles behind the horizons would transfer the heat from
the hotter inner horizon to the colder outer horizon, that takes place due to
the Hawking radiation outside the outer horizon. Therefore, the temperature
of outer horizon will increase during the Hawking radiation. For the case of
Schwarzschild black hole, the increase of outer horizon temperature at fixed
infinite temperature of inner horizon contracted to the singularity means the
decrease of black hole mass, i.e. the evaporation of black hole, since
$T_+^{-1}=8\pi M_\textsc{bh}$. Evidently, that is the infinitely hot
singularity causes the evaporation of Schwarzschild black hole, and that was
our constraint to hold the spacetime structure fixed in order to derive the
Hawking radiation.

However, in thermodynamics, two free thermal systems at different
temperatures will tend to equaling their temperatures due to the heat
transfer between them. Therefore, \textsl{the black hole with different
temperatures of its horizons is thermodynamically instable.} It will tend to
a state with the common temperature of two horizons, as the thermodynamics
claims. Such the process involves the change of spacetime structure of black
hole, i.e. the non-diagonal transition for the gravitational field, of
course. The transition has to result in the emission of gravitational quanta.
The gravitational transition could be accompanied with the radiation of
non-gravitational field, too.

Further we will derive the thermal stability criteria for the Kerr black hole
and calculate the mass of stable black hole in the final state by the
gravitational transition from the initial state of given spacetime structure.

\section{Argumentation and results} The condition of thermodynamical
stability of black hole means that the temperatures of two horizons are equal
to each other. The difference of temperatures is expressed in terms of
horizon areas,
\begin{equation}\label{areas}
    T_- -T_+=\frac{1}{\sqrt{4\pi}}\,
    \frac{(\mathcal A_+-\mathcal A_-)^2}{\mathcal A_+\mathcal A_-}\,
    \frac{1}{\sqrt{\mathcal A_++\mathcal A_-}},
\end{equation}
hence, the stability, i.e. $T_--T_+=0$, certainly gives the following generic
constraint:
\begin{equation}\label{=}
    \mathcal{A}_+=\mathcal{A}_-.
\end{equation}
Thus, any non-extremal black hole will transfer to the extremal one in order
to reach the thermodynamical stability.

However, the constraint of thermal stability in (\ref{=}) does not fix the
value of extremal area $\mathcal{A}_e=\mathcal{A}_+=\mathcal{A}_-$. In order
to find $\mathcal{A}_e$ for the given initial state of black hole, we should
specify the mechanism of transition. Let us consider the transition, when the
gravitational fields are emitted, while virtual particles do not produce the
Hawking radiation.

Trajectories of virtual particles, confined behind the horizons, have been
considered in \cite{KisPRD}. For instance, the spacetime interval of virtual
massive particle in the Schwarzschild background with gravitational radius
$r_g$ is given by
\begin{equation}\label{interval}
    ds^2=\frac{r_c}{r_g}\,\frac{r}{r_c-r}\,dr^2>0,
\end{equation}
where $r_c\leqslant r_g$ is the maximal distance from the singularity, i.e.
$r\leqslant r_c$. We have found that such the trajectories periodically
evolve with the imaginary time for the outer distant observer. This fact
means that virtual particles form the thermal ensemble, and the period
defines the inverse temperature of system\footnote{A similar picture is valid
for the Kerr black hole: we can specify appropriate trajectories of virtual
particles confined between inner and outer horizons. The difference is
reduced to the introduction of two inverse temperatures, which ratio can be
consistently fixed in order to preserve the periodicity behind \textit{two}
horizons.}, $\beta=1/T_\textsc{bh}$. The periodicity determines the discrete
spectrum of $r_c$. The ground state corresponds to $r_c=r_g$.

In classical approximation, the logarithm of partition function is given by
the sum of euclidian actions for the virtual particles per the period. For
the ground state, we find
\begin{equation}\label{ln Z}
    \ln Z=-\beta\,\frac12\sum m,
\end{equation}
where the sum of masses $m$ for the virtual particles is fixed by the
following condition: the black hole mass is equal to the energy of thermal
system, viz. straightforwardly $M_\textsc{bh}=-\partial\ln Z/\partial \beta$.
Then,
\begin{equation}\label{free}
    \frac12\sum m=M_\textsc{bh}-TS,
\end{equation}
which equals the Helmholtz free energy $\mathcal F$.

Notice that eqs. (\ref{ln Z}) and (\ref{free}) is valid for the ground state
of Kerr black hole\footnote{For ground level of massive particle moving at
the equator trajectory, the interval is given by the following expression:
$ds^2={r^2}\,dr^2/{(r_+-r)(r-r_-)},$ in consistency with (\ref{interval}) in
limits of $r_-\to 0,$ $r_c\to r_g\to r_+$.}, too \cite{KisPRD}. Moreover, the
product of temperature to the entropy is invariant with respect to the
horizon prescription, $T_+S_+=T_-S_-.$ Thus, the free energy ascribed to the
black hole is irrelative to the choice of inner or outer horizon.

If the transition between black holes is determined by the emission of
gravitational quanta, then the sum of masses of virtual particles behind the
horizons should be constant, since the process should be given by the
diagonal matrix element of transition for these particles, i.e.
\begin{equation}\label{const}
    \sum m=\mathrm{const.}
\end{equation}
Condition (\ref{const}) means that the free energy of black hole $\mathcal F$
is conserved during the non-diagonal gravitational transition changing the
spacetime structure. In the final state we get the extremal black hole with
the horizon temperature equal to zero,  $T_e=0$, hence, $\mathcal F=M_e$ and
the mass of black hole after the transition is given by the following
expression:
\begin{equation}\label{extreme}
    M_e=M_\textsc{bh}-\frac14\,T_\textsc{bh}\mathcal{A}_\textsc{bh}.
\end{equation}
For instance, if the initial state is the Schwarzschild black hole possessing
$J=0$, then $T_\textsc{bh}=1/8\pi M_\textsc{bh}$, while the horizon area
equals $A_\textsc{bh}=16\pi M_\textsc{bh}^2$, and hence,
\begin{equation}\label{schwarz}
    M_e=\frac12\,M_\textsc{bh}\Big|_{J=0}.
\end{equation}
Remember, that the orbital momentum of extremal black hole equals $J=M_e^2$.

\textit{In addition,} let us comment on the condition of ground level for the
virtual particles used above in our consideration. In \cite{KisHaw} we have
shown, that the black hole emits Hawking radiation until the virtual
particles occupy excited levels, while the radiation stops, if all particles
fall to the ground level. In this sense, our argumentation above becomes more
rigorous, since at the ground level the Hawking radiation is certainly
switched off, and the non-diagonal gravitational processes provide the only
mechanism of transitions between black holes.

Next, as concerns for the quantum spectrum of black holes, usually one puts
$M^2_\textsc{bh}\big|_{J=0}=n/2$ with $n$ being large integer
\cite{BarKust,Bekenstein-2,BM,Kis-spectrum}. Therefore, the transition to
extremal black hole in accordance with (\ref{schwarz}) would give $J=n/8$,
that is not half-integer or integer at arbitrary integer $n$. It means, that
the transition would result in a non-extremal black hole, say, with an
acceptable value of $J<n/8$, so that some particles will be probably excited
from the ground level. Further, the excitations will decay due to the Hawking
radiation, which will decreases the mass of black hole to reach the limit of
extremal black hole at given $J$, for instance. Similar notes could be done
for the case of initial state given by a non-extremal black hole at $J\neq
0$, of course.

Finally, our consideration suggests that all candidates to inactive black
holes, observed in astronomy, are extremal. As for active black hole
candidates, i.e. those of interacting with the ordinary matter, which
surrounds the black hole, we suppose that the interaction is not related with
the transitions described in the paper, since the interaction changes the
mass of extremal black hole due to the accretion of matter or by emission of
matter. The accretion of matter by the extremal black hole with a given
orbital momentum $J$ has to conserve the thermodynamic stability, i.e. the
extremal structure. Therefore, an accreted orbital momentum $\delta J$ has to
adjust an appropriate amount of accreted mass $\delta M^2=\delta J$, but if
the real amount of mass $\delta^\prime M^2$ producing the accretion of
orbital momentum exceeds the adjusted value, then the difference $\Delta
M^2=\delta^\prime M^2-\delta M^2$ has to be ejected by the black hole at the
conserved orbital momentum $\tilde J=J+\delta J$. To our opinion, such the
ejection prefers for the balanced matter jets in opposite directions along
the axis of black hole rotation.

\section{Conclusion} The non-extremal rotating black hole is
thermodynamically instable because its two horizons have different
temperatures, while the heat transfer between them will tend to equalize
these temperatures due to transitions changing the spacetime structure. The
stable state is the extremal black hole having single horizon at zero
temperature. Supposing the transition caused by the emission of gravitational
quanta, we have calculated the mass of extremal black hole in the final
state. The process inherently differers from the evaporation of black hole
due to the Hawking radiation.

\ack

This work is partially supported by the grant of Russian Foundation for Basic
Research, \# 10-02-00061.

\end{document}